# Market-based vs. Price-based Microgrid Optimal Scheduling

Sina Parhizi, *Student Member, IEEE*, Amin Khodaei, *Senior Member, IEEE*, and Mohammad Shahidehpour, *Fellow, IEEE*

***Abstract*—An optimal scheduling model for a microgrid participating in the electricity distribution market in interaction with a Distribution Market Operator (DMO) is proposed in this paper. The DMO administers the established electricity market in the distribution level, sets electricity prices, determines the amount of the power exchange among market participants, and interacts with the Independent System Operator (ISO). Considering a predetermined main grid power transfer to the microgrid, the microgrid scheduling problem will aim at balancing the power supply and demand while taking financial objectives into account. Numerical simulations exhibit the application and the effectiveness of the proposed market-based microgrid scheduling model and further investigate merits over a price-based scheme.**

## NOMENCLATURE

***Indices:***

| | |
|---|---|
| $b$ | Index for buses. |
| $i$ | Index for DERs. |
| $j$ | Index for segments of the load bids. |
| $l$ | Index for transmission lines. |
| $t$ | Index for hours. |
| $m$ | Index for microgrids. |
| $r$ | Superscript for responsive loads. |
| $f$ | Superscript for fixed loads. |

***Parameters and functions:***

| | |
|---|---|
| $B$ | Components of the bus-to-line incidence matrix. |
| $c$ | Marginal cost of dispatchable units. |
| $D^f$ | Total fixed load of all microgrids in the distribution network. |
| $DR$ | Ramp down rate. |
| $F(P,I)$ | Generation cost of the dispatchable unit. |
| $PL_l^{\max}$ | Line flow limit. |
| $UR$ | Ramp up rate. |
| $U$ | Islanding indicator (1 when grid-connected, 0 when islanded). |
| $v$ | Penalty for scheduled power deviation. |
| $\upsilon$ | Value of lost load. |
| $x$ | Line impedance |
| $\rho(P)$ | Cost function of generation units submitted to the ISO. |
| $\lambda(D)$ | Consumption benefit of aggregated loads. |

***Variables:***

| | |
|---|---|
| $d$ | Load demand. |
| $D$ | The demand awarded from the ISO to the DMO. |
| $PD^M$ | Assigned demand to microgrids by the DMO. |
| $DX$ | The amount of load awarded to each segment of the bid |
| $DX^{\max}$ | Segment in the variable load bid of the microgrid. |
| $I$ | Commitment state of dispatchable unit (1 when committed, 0 otherwise). |
| $LS$ | Load curtailment. |
| $P$ | DER output power. |
| $P^M$ | Scheduled power transfer from the DMO to the microgrid. |
| $PL$ | Line flow. |
| $\Delta P^M$ | Power transfer deviation. |
| $\Delta P^{M+}$ | Positive power transfer deviation. |
| $\delta$ | Power transfer deviation indicator (1 when deviation is positive, 0 otherwise). |
| $\theta$ | Bus angle. |

S. Parhizi and A. Khodaei are with the Department of Electrical and Computer Engineering, University of Denver, Denver, CO 80210 USA (emails: sina.parhizi@du.edu; amin.khodaei@du.edu). M. Shahidehpour is with the Department of Electrical and Computer Engineering, Illinois Institute of Technology, Chicago, IL 60616 USA (email: ms@iit.edu).

## I. INTRODUCTION

MICROGRIDS, which were primarily introduced to facilitate the integration of distributed energy resources (DERs) in distribution grids, provide significant benefits to consumers and the system as a whole, which include but are not limited to, improving system reliability and resiliency, providing local intelligence to the customer side, reduction in greenhouse gas emissions, and reducing the need for expanding transmission and distribution facilities as a result of generation-load proximity. In addition, the microgrid capability to be operated as a single controllable entity, which is enabled through application of dispatchable DERs and flexible loads, allows for an active participation in a variety of demand response programs in response to economic and emergency incentives [1]–[9].

An efficient operation and control is one of the most challenging aspects in managing microgrids. The microgrid control is commonly performed in three hierarchical levels, including primary, secondary, and tertiary [10], [11]. The first two control levels deal with droop control and frequency/voltage adjustment and restoration when there is a change in the microgrid load and/or generation as well as in islanding transitions. The third control level schedules microgrid components and determines the interactions with the main grid while taking economy and reliability aspects into consideration. Microgrid scheduling problem aims to minimize the operation costs of local DERs, as well as the energy exchange with the main grid, to supply forecasted local loads in a certain period of time (typically one day). This problem is subject to a variety of operational constraints, such as power balance and DER limitations, and is performed by the microgrid controller. Extensive discussions are available in the literature on the architecture of the microgrid controller, including decentralized [12], [13] centralized [14]–[16], and hybrid microgrids [17], [18], and also on the methods to solve the scheduling problem with a primary focus on accurate component modeling and uncertainty consideration, such as deterministic methods [19]–[21], stochastic programming [22], [23], chance-constrained [24], and robust optimization [12]. Additional discussions can be found in the literature, solving the problem using multi-agent systems [25], or benefiting from heuristic methods such as particle swarm optimization [26].

Increasing demand-side elasticity and active participation of loads in the power system, commonly in response to electricity price variations, is highly stressed to operate the system more efficiently and to avoid high price spikes caused by inelastic loads [27]. Microgrids allow an efficient integration and control of large penetration of responsive loads which would further increase the demand-side elasticity. Moreover, distributed generators (DGs) and energy storage support a relatively fast and highly controllable load. However, these resources are typically scheduled based on a price-based scheduling model, i.e., the microgrid controller determines the least-cost schedule of available DERs and loads, as well as the main grid power transfer, based on the day-ahead market prices (which are forecasted by the microgrid or the electric utility). Under this scheme, the utility forecasts an estimate of the microgrids' loads in its service territory and submits it to the system operator. Once the price of electricity is determined, through the wholesale market, the utility sends the actual prices to microgrids. Although it might seem efficient, this approach has the potential to cause several drawbacks when the microgrid penetration in distribution network is high, including but not limited to shifting the peak hours. This approach is prone to cause new peaks since there is a high probability that microgrids follow a different schedule compared to the one forecasted by the utility once actual prices are received, considering that the power demand in responsive loads is inversely proportional to electricity prices. The increase in the number of entities with responsive loads operated based on price-based scheme would intensify this issue. In other words, setting the price centrally by the system operator and sending it to microgrids, so they can accordingly schedule their resources, can potentially result in significant uncertainty in the system load profile. The increased penetration of DERs and microgrids would also make it more challenging to ensure distribution system reliability [28].

The concept of aggregators was one of the ideas that was proposed to address these issues. Aggregators discussions can be found in [29], where it is proposed to iteratively collect power generated by microgrids, sell this power to the main grid, and accordingly gain profit via a price-based scheduling. In [30] an aggregator for electric vehicles with fixed energy cost is proposed. The study in [31] proposes a framework for interactions between the customers in a distribution system as well as the main grid, while [32] proposes an entity between the market operator and customers that compensates the aggregators for the services they provide. A coupon incentive-based demand response model is further proposed in [33] enabling customers to increase their flexibility and lower their costs. The proposed model in [34] enables a demand response aggregator to participate in the electricity market, considering market price to be constant. It is further applied to microgrids in [35].

The aforementioned drawbacks, combined with the enhanced complexity in managing a large number of microgrids in a foreseeable future, has made the case for developing new methodologies for the system operation and utility ratemaking in presence of microgrids. The concept of a Distribution System Operator (DSO) is recently proposed as an entity which is hosted in the distribution network to manage interaction of microgrids with the main grid. In line with the ongoing trend in proposing electricity markets in distribution networks, this paper proposes a market-based microgrid optimal scheduling model to address the aforementioned problems and increase microgrid-integrated distribution system efficiency and social welfare. Considering that a DSO offers both grid and market functionalities, this paper only focuses on the market operation and provides discussions under a Distribution Market Operator (DMO) concept. The DMO can be considered as the distribution level equivalent of the ISO, which is responsible for managing the electricity market and scheduling power transfers to achieve the optimal operation in the distribution network [36]. Considering limited research studies on the viability of distribution markets, this paper aims to: (i) discuss the necessity of the DMO in future power grids and identify interactions with connected market players, (ii) formulate the three levels of the market structure, i.e., ISO, utility, and customer levels, to provide an insight on the data exchange and involved optimization problems in these levels, (iii) develop an analytical model for the market-based microgrid scheduling, and formulate the problem using mixed-integer linear programming, and (iv) use the developed comprehensive model to enable comparisons between price-based and market-based scheduling schemes from microgrids and system perspectives.

The rest of the paper is organized as follows: Section II elaborates on the necessity of system operators and market mechanism at the distribution level. Section III outlines the DMO model for distribution networks and compares it with the traditional architecture. Section IV formulates the market-based microgrid optimal scheduling model along with models for the ISO and the DMO to enable a comprehensive study. Section V presents the numerical results, comparisons, and discussions. Finally, Section VI concludes the paper.

## II. Distribution System Operators

It may be discussed that a highly accurate load estimation would resolve the mentioned issues, in which the system operators would have a fairly accurate idea of load variations. However, it should be noted that necessity of the DSO is not limited to improving predictability. The DSO provides the local resilience capability [28] and reduces dependence on the ISO for providing balancing services, so the distribution system can maintain its service when the rest of the system is in abnormal condition [37]. It could also manage the energy transactions happening between the DERs and loads within the distribution system; demand for this service would grow as the number of such transactions increase [28]. New York Reforming The Energy Vision (REV) asserts that in order to "create a more robust retail market" it is necessary to provide market operations and grid operations at the distribution level [38]. Easing complexity of direct scheduling of responsive loads and DERs in the wholesale market, solving scalability issues and providing ancillary services are among other beneficial functionalities that the DSO can provide to the distribution system [37]–[39]. On the other hand, the ISOs may not have control over the demand side assets, so those assets need to have the capability to provide reserve and flexibility services to handle variable resources [40].

In [41] a price-based simultaneous operation of microgrids and the Distribution Network Operator (DNO) is proposed. In New York, the new concept of Distributed System Platform Provider (DSPP) is introduced as part of the Reforming the Energy Vision program [38] where the transformation of existing utility operations to integrate high penetration of microgrids and DERs is discussed. DSPPs can be formed as new entities or be part of the currently existing electric utilities. An independent DSPP would be able to set up a universal market environment instead of one for each utility. It would also be less suspected of exercising market power. A utility-affiliated DSPP, however, would be able to perform several functionalities currently possessed by electric utilities without necessitating additional investments. In California, the state public utilities commission has ruled to establish regulations to guide investor-owned electric utilities in developing their Distribution Resources Plan proposals. Studies in [28], [42] provide a framework for this ruling, defined as a DSO, which is in charge of operation of local distribution area and providing distribution services. The DSO is further responsible to provide forecasting and measurement to the ISO and manage power flow across the distribution system. It is also suggested that the DSO adopt further roles such as coordination of dispersed units in the distribution network and providing an aggregate bid to the ISO. The study in [40] proposes a DSO as an ISO for the distribution network, which is responsible for balancing supply and demand at the distribution level, linking wholesale and retail market agents, and linking the ISO to the demand side. As opposed to the European definition of the DSO, the proposed entity in [40] interacts directly with the ISO. The study further presents a spectrum of different levels of the DSO autonomy in operating the distribution system and the degree of ISO's control over it. From the least autonomy to the most autonomy, this spectrum entails the DSO to be able to perform the forecasting and send it to the ISO, be responsible for balancing the supply and demand, be able to receive offers from DERs, aggregate them and bid into the wholesale market, and eventually be able to control the retail market so that various DERs can have transactions not only with the DSO but among themselves. In [43], an independent distribution system operator (IDSO) is proposed to be responsible for distribution grid operation, while grid ownership remains in the hands of utilities. The IDSO is envisioned to provide market mechanisms in the distribution system, enable open access, and ensure safe and reliable electricity service. The IDSO will reduce the operation burden on utilities and determine the true value of resources more objectively. The necessity of distribution markets in integrating proactive customers has been emphasized in the literature [44]-[46].

## III. Model Outline

The discussed DMO in this paper is a platform that enables market activities for end-use customers, coordinates with the utility to improve grid operations, and interacts with the ISO to determine demand bid awards. The DMO will further facilitate establishing a competitive electricity market in the distribution network to exchange energy and grid services with customers, and expedite a more widespread integration of DERs from a system operator's perspective by addressing prevailing integration challenges. Fig. 1 depicts the interactions of different players in the market in presence of the DMO via three levels of ISO, Utility, and Customer.

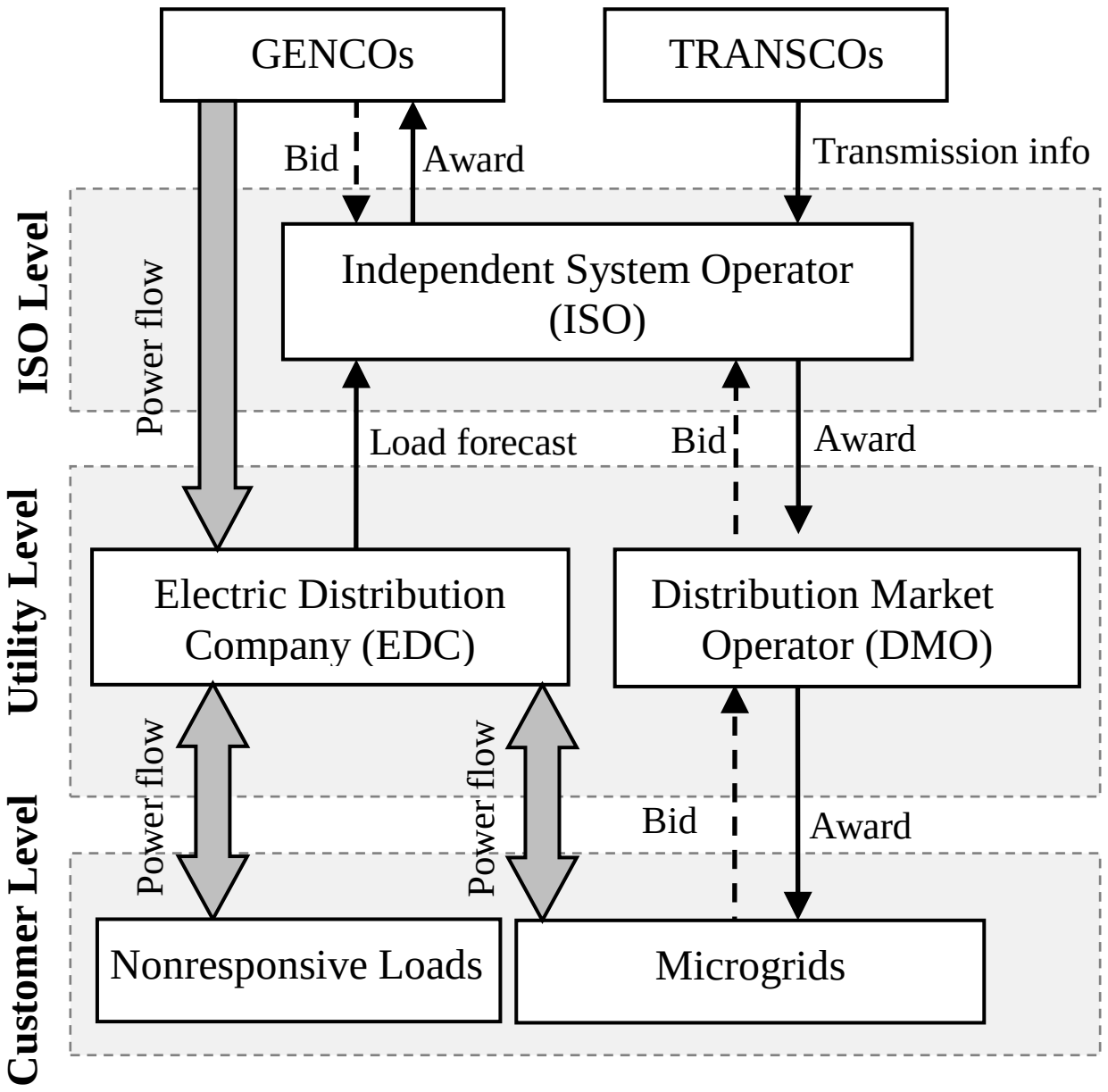

Fig.1 Microgrid market participation through the proposed DMO

Two major responsibilities of the DMO within this structure are: 1) To receive demand bids from the microgrids (and other responsive loads if any), combine them, and offer an aggregated bid to the ISO, and 2) To receive the day-ahead schedule from the ISO, solve a resource scheduling problem for its service territory, and subsequently determine microgrids' shares from the awarded power. Microgrids would submit their bids to the DMO (in the form of monotonically decreasing demand bids) and later be notified by the DMO on the amount of awarded power (henceforth referred to as the assigned power). Other responsive loads can be considered at the Customer level without loss of generality. The main grid power transfer to the microgrid would be the amount of power assigned to the microgrid by the DMO, hence it would be known to the system operator in advance and therefore eliminate the uncertainties caused by microgrids to a large extent. Once the main grid power transfer is reported to the microgrid, for the 24 hours of the next day, the microgrid would solve a market-based scheduling problem to optimally schedule its DERs and loads. Only private microgrids are considered in this paper as there could be some regulatory barriers in market participation of utility-owned microgrids.

This framework offers several advantages:

- The microgrid demand is set by the DMO and known with certainty on a day-ahead basis. This will lead to manageable peak demands and increased operational reliability and efficiency. Microgrids will have the capability to deviate from the assigned power (as it will be further discussed in this paper), however it will be at the expense of paying a penalty, hence potential deviations would be minimal.
- The microgrid can exchange power with the main grid and act as a player in the electricity market. The DMO would serve as an intermediate entity between the ISO and microgrids that facilitates microgrids market participation and coordinates the microgrids with the main grid to minimize the risks posed by microgrids operational uncertainties.
- Establishing the DMO is beneficial to the ISO as it allows a significant reduction in the required communication infrastructure among microgrids and the ISO.
- The DMO can be formed as a new entity or be part of the currently existing electric utilities. An independent DMO would be able to set up a universal market environment instead of one for each utility. It would also be less suspected of exercising market power. On the other hand, a utility-affiliated DMO would be able to perform several functionalities currently possessed by electric utilities without necessitating additional investments.

Implementation of the DMO would fix the aforementioned problems that utilities face when they integrate microgrids. However, in order for the proposed framework to work reliably, it is necessary that the microgrid controller schedules its resources based on the assigned power transfer, considering that microgrid controller seeks the least-cost schedule of local resources. It will be assumed that deviations from the assigned value will be penalized based on the market price or a relatively larger value that can effectively prevent and/or reduce deviations. In this paper, it is assumed that the penalty will be applied when the deviation is positive, i.e., the microgrid's scheduled power is larger than the assigned power by the DMO, or in other words, when the microgrid appears as a larger load compared to the assigned power by the DMO. Negative deviation will not be penalized in the proposed model as the microgrid helps with reducing load (increasing generation) in the distribution network. This issue is further investigated in numerical simulations.

In the price-based scheduling method, the ISO receives load forecasts from Electric Distribution Companies (EDC) and determines the day-ahead unit schedules by solving a security-constrained unit commitment (SCUC) problem. The ISO will also determine locational marginal prices (LMPs) which will be further used by microgrids for scheduling purposes. In the price-based method, contrary to the market-based method, there is no need for microgrids to offer bids and participate in the market, and moreover, the main grid power transfer will be determined via a local cost minimization problem rather than being determined by the DMO via a market mechanism. In either method, however, the microgrid needs to determine the optimal schedule of local DERs and loads to address its energy needs and ensure a reliable supply of local loads.

## IV. Market-Based Microgrid Optimal Scheduling Problem Formulation

The discussed three levels of the market structure are formulated in the following to provide an insight on the data exchange, represent optimization problems involved in different levels, and further enable numerical studies on microgrids scheduling.

### A. Microgrid

Microgrids determine the least-cost day-ahead schedule of their loads, dispatchable generation units, and energy storage considering a known profile for the main grid power transfer, which is determined and assigned by the DMO, over the scheduling horizon. Each microgrid $m$ solves the proposed market-based optimal scheduling problem (1)-(10):

$$\min \sum_t \sum_i [F_{im}(P_{imt}, I_{imt}) + \upsilon_m LS_{mt} + \nu_m \Delta P_{mt}^{M+}] \tag{1}$$

$$P_{mt}^M + \sum_i P_{imt} + LS_{mt} = d_{mt} \qquad \forall t \tag{2}$$

$$P_{im}^{\min} I_{imt} \le P_{imt} \le P_{im}^{\max} I_{imt} \qquad \forall t, \forall i \tag{3}$$

$$\sum_t P_{imt} = E_{im} \qquad \forall t \tag{4}$$

$$\sum_t f_{im}(P_{imt}, I_{imt}) \le 0 \qquad \forall i \tag{5}$$

$$-P_m^{M,\max} U_{mt} \le P_{mt}^M \le P_m^{M,\max} U_{mt} \qquad \forall t \tag{6}$$

$$\Delta P_{mt}^M = P_{mt}^M - PD_{mt}^M \qquad \forall t \tag{7}$$

$$-P^{M,\max} \delta_{mt} \le \Delta P_{mt}^{M+} \le P^{M,\max} \delta_{mt} \qquad \forall t \tag{8}$$

$$-P^{M,\max}(1-\delta_{mt}) \le \Delta P_{mt}^M - \Delta P_{mt}^{M+} \le P^{M,\max}(1-\delta_{mt}) \qquad \forall t \tag{9}$$

$$-P^{M,\max}(1-\delta_{mt}) + \varepsilon \le \Delta P_{mt}^M \le P^{M,\max} \delta_{mt} \qquad \forall t \tag{10}$$

The three terms in the objective function (1) represent the operation cost, the load curtailment cost, and the deviation cost, respectively. The operation cost is the cost of power production as well as startup and shut down costs of dispatchable units. The load curtailment cost is defined as the value of lost load times the amount of load curtailment. The value of lost load is assumed as an opportunity cost based on the cost that the consumer is willing to pay to have reliable uninterrupted service. It is commonly used as a measure of load criticality [47]. The deviation cost is the penalty imposed on the microgrid in case the microgrid schedule deviates from the power transfer assigned by the DMO. The objective is subject to a set of operational constraints (2)-(10). The power balance equation (2) ensures that the sum of the main grid power plus the locally generated power from DERs matches the total load, while load curtailment variable is added to ensure that the power balance is satisfied at all times. Nondispatchable generation and fixed load values are assumed to be forecasted with acceptable accuracy and are treated as uncontrollable parameters. There of course would be uncertainties associated with possible forecast errors, which will be studied in a follow-on work. All operational constraints associated with DERs and loads are formulated using three general constraints (3)-(5), respectively representing power constraints, energy constraints, and time-coupling constraints. Power constraints (3) account for power capacity limits, such as dispatchable generation minimum/maximum capacity limits, energy storage minimum/ maximum charge/discharge power, and flexible load minimum/maximum capacity limits. Energy constraints (4) account for energy characteristics of a specific DER or load, such as energy storage state of charge limit and flexible load required energy in a cycle. Time-coupling constraints (5) represent any constraint that links variables in two or more scheduling hours, such as dispatchable units ramp up/down rates and minimum on/off times, energy storage rate and profile of charge/discharge, and adjustable loads minimum operating time and load pickup/drop rates. Using these constraints, any type of DER and load can be efficiently modeled. A detailed modeling of microgrid DERs and loads can be found in [48]. The main grid power transfer is restricted by its associated limits (imposed by the capacity of the line connecting the microgrid to the main grid) in (6). The islanding is modeled using a binary islanding indicator $U$ which would zero out the main grid power transfer when 0. The main grid power deviation to be penalized in the objective is determined in (7)-(10). Constraint (7) calculates the deviation by subtracting the scheduled power via the optimal scheduling, $P^M$, by the assigned power from the DMO, $PD^M$. Constraints (8)-(10) determine the penalty if the calculated deviation is positive. An auxiliary binary variable $\delta$ is used for this purpose. When $\delta$=0 the power transfer to be penalized is zero, i.e., the scheduled power is less than the assigned power. However, when $\delta$=1 the power transfer to be penalized is equal to the positive deviation calculated in (7).

## B. DMO

The DMO seeks two objectives: first, to combine individual bids received from microgrids in its territory to create an aggregated bid and accordingly send the aggregated bid to the ISO to participate in the energy market; second, to disaggregate the awarded quantity by the ISO to individual microgrids in accordance with their respective bids. These two tasks are discussed in the following:

*Bid aggregation:* Fig. 2 depicts a typical demand bid curve submitted by a microgrid to the DMO at a specific hour t. The bid consists of fixed and variable parts. The fixed part shows the microgrid nonresponsive load which must be fully supplied under normal operation conditions and cannot be altered. The variable part, on the other hand, shows the microgrid flexibility in reducing its consumption from its total load. It consists of several segments. The reduction in consumption can be achieved either via load curtailment or local DER generation. The DMO combines the individual microgrid bids and obtains an aggregated bid to be sent to the ISO. The fixed loads are collectively added to obtain the total fixed load in the DMO service territory (11).

$$D_t^f = \sum_m d_{mt}^f \qquad \forall t \qquad (11)$$

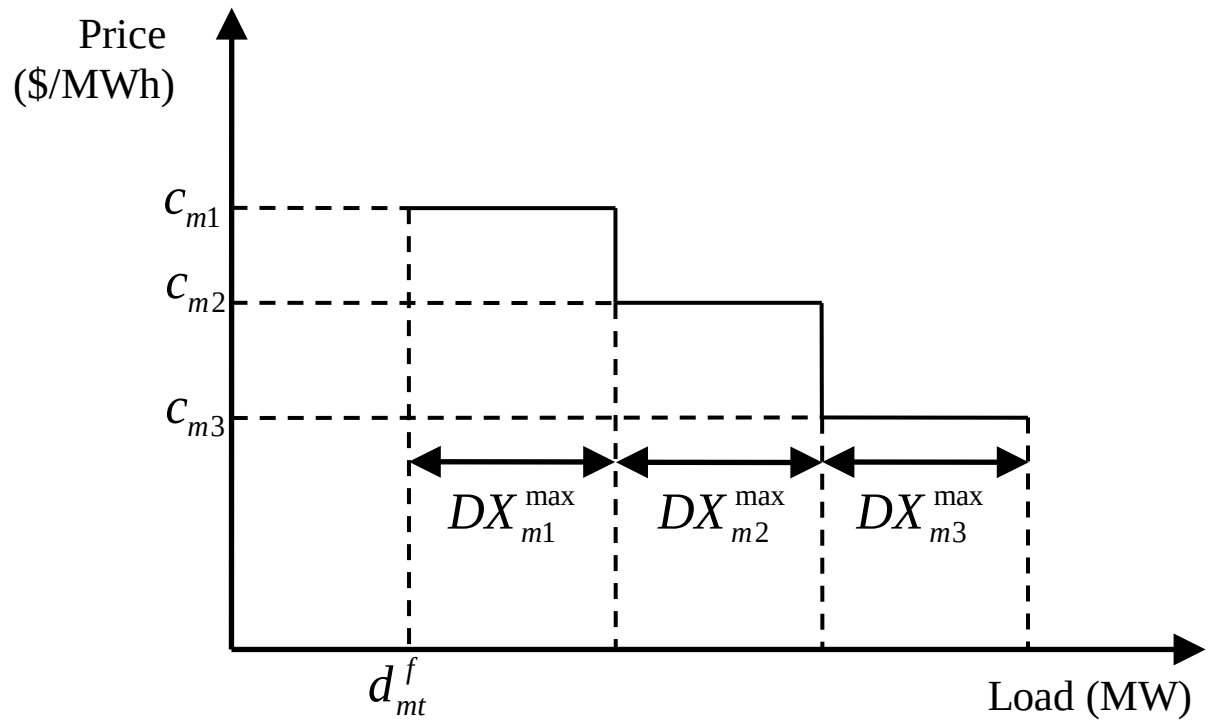

Fig. 2 Demand bid curve for microgrid *m* with a three-segment bid.

*Quantity disaggregation:* Once the ISO determines the awarded power to the DMO, the DMO disaggregates the power to microgrids in its service territory. The DMO maximizes the objective function (12) by determining the optimal allocated power to each microgrid based on the submitted bids.

$$\max \sum_t \sum_m \sum_j c_{mj} DX_{mjt} \qquad (12)$$

$$DX_{mjt} \le DX_{mj}^{\max} \qquad \forall m, \forall t, \forall j \qquad (13)$$

$$d_{mt}^r = \sum_j DX_{mjt} \qquad \forall m, \forall t \qquad (14)$$

$$d_{mt}^f + d_{mt}^r = PD_{mt}^M \qquad \forall m, \forall t \qquad (15)$$

$$\sum_m PD_{mt}^M = D_{bt} \qquad \forall t \qquad (16)$$

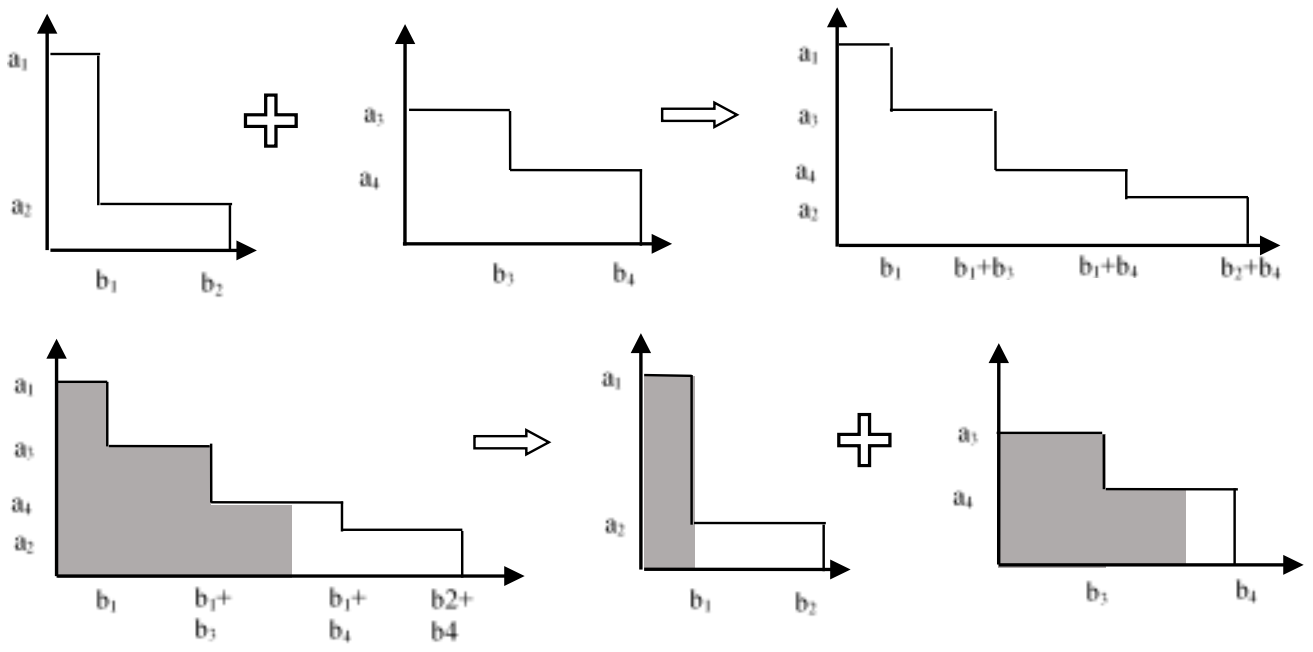

Fig. 3 An example of DMO aggregating two submitted bids (top), and disaggregating awarded power (bottom). Vertical and horizontal axes show price and load, respectively.

Constraint (13) guarantees that each segment of load is limited by its maximum. The total responsive demand for each microgrid is the sum of the loads dispatched to each associated segment (14). The awarded load is calculated as the summation of the fixed and responsive loads (15), and accordingly, the amount of power flow from the ISO to the DMO as the summation of the awarded loads is set by (16) as the total load dispatched to all load segments is equal to the assigned power by the ISO. Fig. 3 provides a graphical representation of the bid aggregation and quantity disaggregation by the DMO. The distribution line limits in this model are assumed to be adequately large to handle any power transfer without causing congestion in the distribution network. Additional constraints, however, can be simply added to the model, including but not limited to distribution line power flow and limits, ramp rate constraints, etc. Another important constraint that can be considered is the load shifting capability of microgrids. Modeling the load shifting would require the inclusion of time-coupling constraints among hourly bids. This topic is addressed for ISOs in previous work of authors [49]. For market-based microgrid optimal scheduling problem, however, load shifting will be considered in a follow-on research of this work.

### C. *ISO*

The ISO receives the generation and transmission information from GENCOs and TRANSCOs, and demand bids from DMOs, solves the SCUC problem to determine units schedule followed by a security-constrained optimal power flow to determine unit dispatch, line flow, and LMPs. The ISO's objective, when considering demand bids, will be to maximize the system social welfare, rather than minimizing the total operation cost, as formulated in (17).

$$\max\left\{\sum_t\sum_b \lambda_{bt}(D_{bt}) - \sum_t\sum_i \rho_{it}(P_{it})\right\} \quad (17)$$

$$\sum_{i\in G_b} P_{it} - \sum_{l\in L_b} PL_{lt} = D_{bt} \qquad \forall t, \forall b \quad (18)$$

$$P_i^{\min} I_{it} \le P_{it} \le P_i^{\max} I_{it} \qquad \forall t, \forall i \quad (19)$$

$$\sum_t P_{it} = E_i \qquad \forall t \quad (20)$$

$$\sum_t f_i(P_{it}, I_{it}) \le 0 \qquad \forall t, \forall i \quad (21)$$

$$|PL_{lt}| \le PL_l^{\max} \qquad \forall t, \forall l \quad (22)$$

$$PL_{lt} = \sum_b \frac{B_{lb}\theta_{bt}}{x_l} \qquad \forall t, \forall l \quad (23)$$

The ISO maximizes objective function (17) which is the system social welfare, i.e., consumption payments minus generation costs. This objective is subject to the power balance constraint (18), unit constraints (19)-(21), transmission line limits (22), and transmission line power flow (23). Unit constraints include unit output limits, unit spinning/operating reserve limit, ramp up/down rate limits, min up/down time limits, fuel limits, and emission limits. Details of the SCUC model can be found in [49].

## V. Numerical Simulations

The proposed market-based microgrid scheduling model is studied and compared with the price-based scheduling using the IEEE 118-bus standard test system (shown in Fig. 4). A total of 5 microgrids are considered to be connected to bus 60 with a total installed DG capacity of 50 MW which is equal to 51% of the peak load at this bus. The specifications of microgrid DGs are given in Table I. Specifications of adjustable loads, energy storage, and fixed loads are borrowed from [48]. Two cases are considered as follows:

**Case 1:** Price-based microgrid optimal scheduling.
**Case 2:** Market-based microgrid optimal scheduling.

**Case 1:** In this case, the ISO uses the forecasted microgrid loads to clear the market and accordingly determine hourly LMP values. Microgrids individually perform their own scheduling using the LMP values. With microgrids being connected to bus 60, five lines in the system including one of those connected to the bus 60 become congested at peak hours. The total microgrid operation cost is calculated as $74,447. In this case, the actual amount of load at the bus to which microgrids are connected will not match the amount originally considered by the ISO when clearing the wholesale market. If the ISO runs the economic dispatch with the actual microgrids net loads, which would be less than the microgrid load used initially by the ISO to determine the LMPs, the prices would change. This change in LMPs can be considered as a major drawback in the price-based model where there is a mutual and uncontrolled interaction between the calculated LMPs and microgrids net load. Another drawback that needs to be considered is that the mismatch between the initially forecasted load and the actual load, after microgrid optimal scheduling, needs to be addressed by the ISO by redispatching the committed units. The redispatch will potentially result in an economic loss for the system as the new solution will diverge from the already determined optimal dispatch solution.

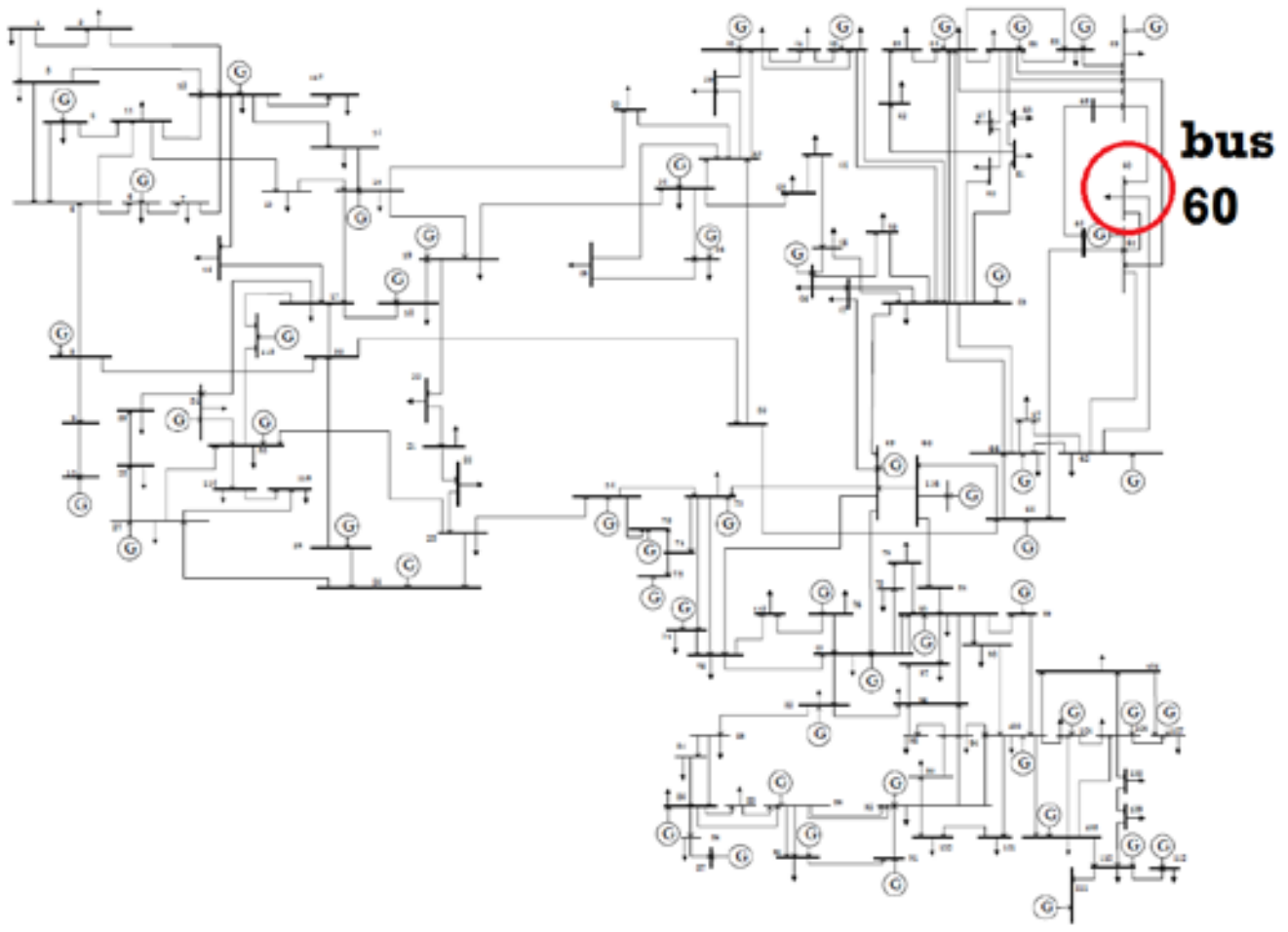

Fig. 4 IEEE 118-bus standard test system.

**Case 2:** The bid each microgrid sends to the DMO is created based on the capacity and marginal costs of its dispatchable DGs. For example, microgrid 2 will have a four-step bid: 1 MW at $70.9/MWh, 1 MW at $59.3/MWh, 3 MW at $37.3/MWh, and 5 MW at $29.1/MWh, as derived from Table I. Using this bid, the demand responsiveness of the microgrid is modeled by local generation of dispatchable DGs. The total microgrid operation cost in this case is $48,568 which shows 34.76% reduction from that of Case 1. Table II shows the committed DGs in each microgrid, in which bold values represent changes from the price-based optimal scheduling solution in Case 1. This table indicates that many DGs committed in Case 1 are not committed in Case 2. In Case 1, microgrid lowers its power transfer as a response to the market price, therefore it has to commit more local resources to

supply loads. DG1 of each microgrid is the most committed unit in both cases, since it has the lowest marginal cost compared to other DGs in the same microgrid.

Fig. 5 depicts the hourly net load at bus 60 to which microgrids are connected. It is observed that during the early hours the values of net load in the two cases are close. This is due to the low price of electricity during early hours, when a large portion of the submitted bid from the DMO is awarded by the ISO, resulting in a power transfer close to the total load of the microgrids. In Case 1, the entire demand is supplied by the main grid for the same reason. At hours 8-24, as the electricity price increases, the microgrids loads are partially supplied by local DGs.

TABLE I
COST CHARACTERISTICS OF DG UNITS

| | Capacity (MW) | | | | |
|---|---|---|---|---|---|
| | MG1 | MG2 | MG3 | MG4 | MG5 |
| DG1 | 4 | 5 | 3 | 4 | 3 |
| DG2 | 3 | 3 | 3 | 3 | 3 |
| DG3 | 2 | 1 | 2 | 2 | 2 |
| DG4 | 1 | 1 | 2 | 1 | 2 |
| | Price ($/MWh) | | | | |
| DG1 | 27.5 | 29.1 | 27.4 | 28.3 | 33.5 |
| DG2 | 43.1 | 37.3 | 38.2 | 35.3 | 41.1 |
| DG3 | 64.3 | 59.3 | 55.2 | 60.3 | 65.5 |
| DG4 | 69.6 | 70.9 | 61.1 | 62.4 | 72.2 |

TABLE II
THE COMMITMENT SCHEDULE OF MICROGRID DGS

| | | 1-24 | | | | | | | | | | | | | | | | | | | | | | | |
|---|---|---|---|---|---|---|---|---|---|---|---|---|---|---|---|---|---|---|---|---|---|---|---|---|---|
| MG 1 | DG1 | **1** | **1** | **1** | **1** | **1** | 0 | **0** | **0** | **0** | **0** | 1 | 1 | 1 | 1 | 1 | 1 | 1 | 1 | 1 | 1 | 1 | 1 | 1 | 1 |
| | DG2 | 0 | 0 | 0 | 0 | 0 | 0 | 0 | 0 | 0 | 0 | 0 | **0** | **0** | **0** | **0** | **0** | **0** | **0** | **0** | **0** | **0** | **0** | 0 | 0 |
| | DG3 | 0 | 0 | 0 | 0 | 0 | 0 | 0 | 0 | 0 | 0 | 0 | 0 | **0** | **0** | **0** | **0** | **0** | **0** | **0** | **0** | **0** | 0 | 0 | 0 |
| | DG4 | 0 | 0 | 0 | 0 | 0 | 0 | 0 | 0 | 0 | 0 | 0 | 0 | **0** | **0** | **0** | **0** | **0** | **0** | **0** | **0** | **0** | 0 | 0 | 0 |
| MG 2 | DG1 | 0 | 0 | 0 | **1** | **1** | **1** | 0 | **0** | **0** | 1 | 1 | 1 | 1 | 1 | 1 | 1 | 1 | 1 | 1 | 1 | 1 | 1 | 1 | 1 |
| | DG2 | 0 | 0 | 0 | 0 | 0 | 0 | 0 | **0** | **0** | **0** | **0** | **0** | **0** | **0** | **0** | **0** | **0** | 1 | 1 | 1 | **0** | **0** | 0 | 0 |
| | DG3 | 0 | 0 | 0 | 0 | 0 | 0 | 0 | 0 | 0 | 0 | 0 | 0 | **0** | **0** | **0** | **0** | **0** | **0** | **0** | **0** | **0** | 0 | 0 | 0 |
| | DG4 | 0 | 0 | 0 | 0 | 0 | 0 | 0 | 0 | 0 | 0 | 0 | 0 | **0** | **0** | **0** | **0** | **0** | **0** | **0** | **0** | **0** | 0 | 0 | 0 |
| MG 3 | DG1 | **1** | **1** | **1** | **1** | **1** | 0 | 0 | **0** | **0** | 1 | 1 | 1 | 1 | 1 | 1 | 1 | 1 | 1 | 1 | 1 | 1 | 1 | 1 | 1 |
| | DG2 | 0 | 0 | 0 | 0 | 0 | 0 | 0 | **0** | **0** | **0** | **0** | **0** | **0** | **0** | **0** | 1 | 1 | 1 | **0** | **0** | **0** | **0** | 0 | 0 |
| | DG3 | 0 | 0 | 0 | 0 | 0 | 0 | 0 | 0 | 0 | 0 | 0 | 0 | **0** | **0** | **0** | **0** | **0** | **0** | **0** | **0** | **0** | 0 | 0 | 0 |
| | DG4 | 0 | 0 | 0 | 0 | 0 | 0 | 0 | 0 | 0 | 0 | 0 | 0 | **0** | **0** | **0** | **0** | **0** | **0** | **0** | **0** | **0** | 0 | 0 | 0 |
| MG 4 | DG1 | 0 | 0 | 0 | 0 | 0 | **1** | **1** | 1 | 1 | 1 | 1 | 1 | 1 | 1 | 1 | 1 | 1 | 1 | 1 | 1 | 1 | 1 | 1 | 1 |
| | DG2 | 0 | 0 | 0 | 0 | 0 | 0 | 0 | **0** | **0** | **0** | **0** | **0** | **0** | **0** | **0** | 1 | 1 | 1 | **0** | **0** | **0** | **0** | **0** | 0 |
| | DG3 | 0 | 0 | 0 | 0 | 0 | 0 | 0 | 0 | 0 | 0 | 0 | 0 | **0** | **0** | **0** | **0** | **0** | **0** | **0** | **0** | **0** | 0 | 0 | 0 |
| | DG4 | 0 | 0 | 0 | 0 | 0 | 0 | 0 | 0 | 0 | 0 | 0 | 0 | **0** | **0** | **0** | **0** | **0** | **0** | **0** | **0** | **0** | 0 | 0 | 0 |
| MG 5 | DG1 | 0 | 0 | 0 | 0 | 0 | 0 | 0 | **0** | **1** | **1** | **1** | **0** | **0** | **0** | **0** | **0** | **0** | 1 | 1 | 1 | **0** | **0** | **0** | 0 |
| | DG2 | 0 | 0 | 0 | 0 | 0 | 0 | 0 | **0** | **0** | **0** | 1 | **1** | **1** | **0** | **0** | **0** | **0** | **0** | **0** | **0** | **0** | **0** | 0 | 0 |
| | DG3 | 0 | 0 | 0 | 0 | 0 | 0 | 0 | 0 | 0 | 0 | 0 | 0 | **0** | **0** | **0** | **0** | **0** | **0** | **0** | **0** | **0** | 0 | 0 | 0 |
| | DG4 | 0 | 0 | 0 | 0 | 0 | 0 | 0 | 0 | 0 | 0 | 0 | 0 | **0** | **0** | **0** | **0** | **0** | **0** | **0** | **0** | **0** | 0 | 0 | 0 |

Fig. 6 depicts the hourly LMP at bus 60, i.e. the electricity price for the power transferred to the DMO. Case 2 represents a significantly lower price in peak and close to peak hours, which accordingly results in fewer DG commitments, as it is more economical to purchase power from the main grid, and a lower operation cost. Accordingly the microgrid net load is increased in this case. Fig. 7 depicts the average LMP of all buses in the system. The values for market-based model are close to or lower than the values for the price-based model except for hours 13 to 22. This result advocates that although the market-based scheduling may result in lower LMPs for microgrids, it may not necessarily reduce the system LMP on other network buses. The total system operation cost is reduced from $1,074,504 in the price-based model to $1,009,734 in the market-based model. To identify the changes in values/trends of LMPs, when such a market is available at all network buses, is worth further investigation.

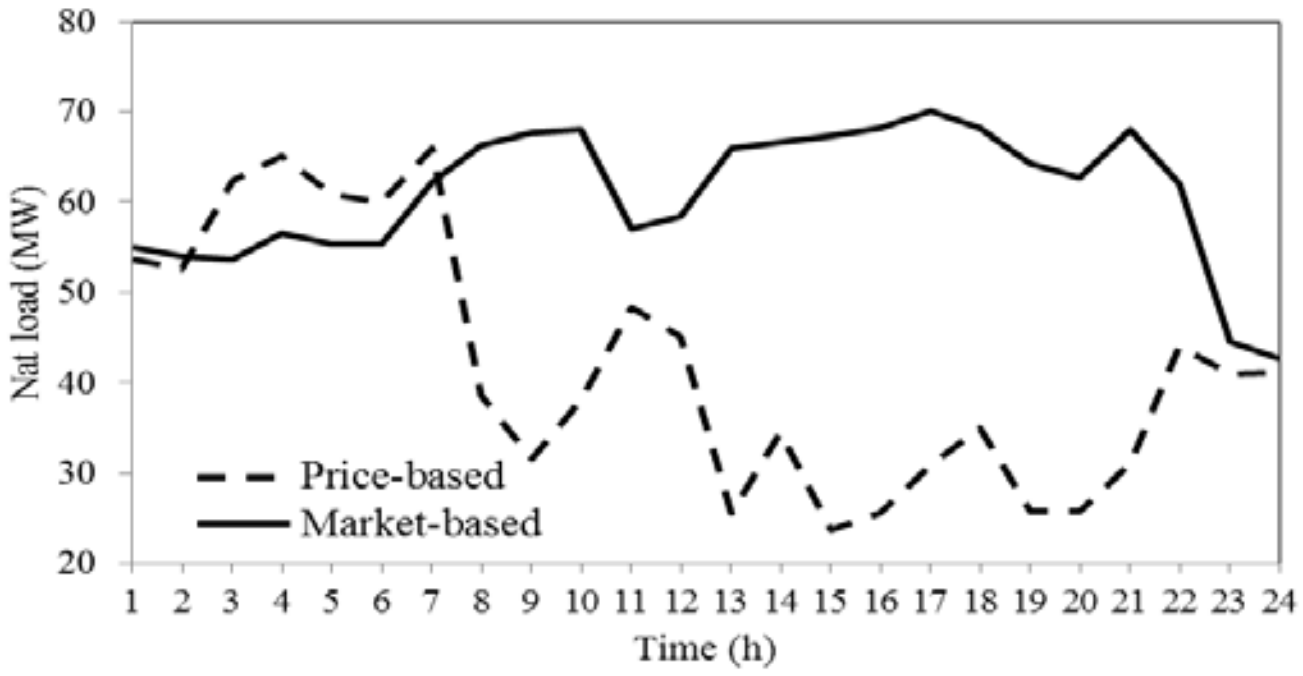

Fig. 5 Net load at bus 60

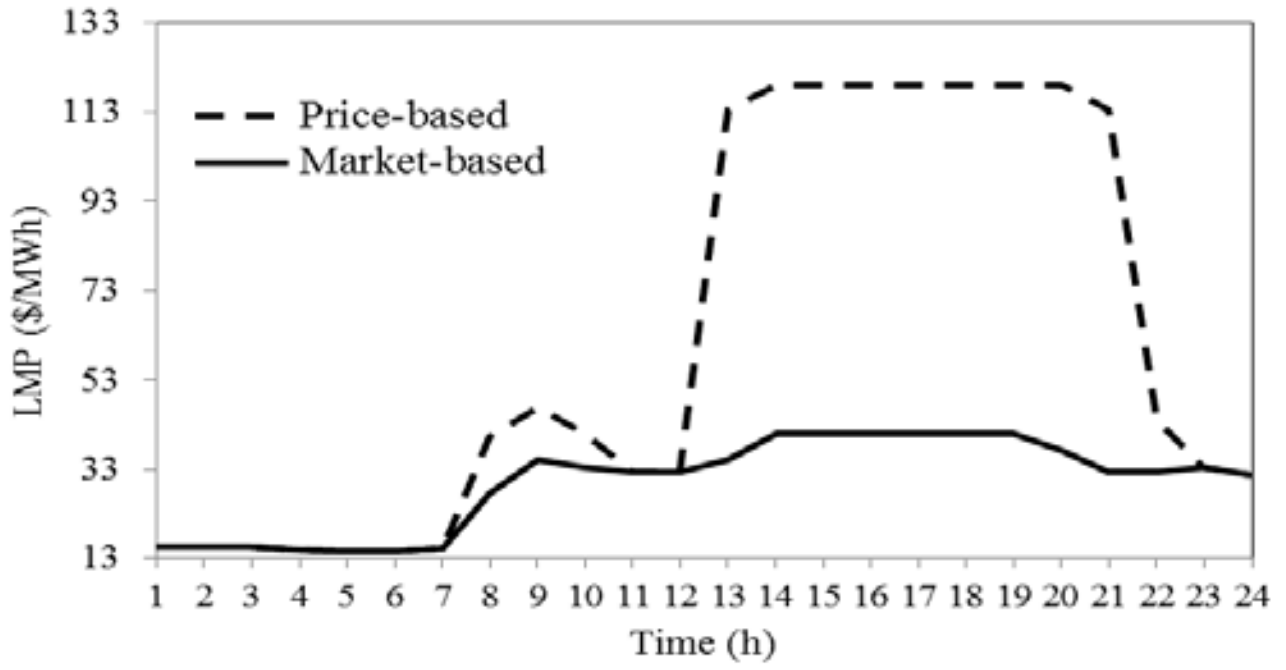

Fig. 6 LMP at bus 60

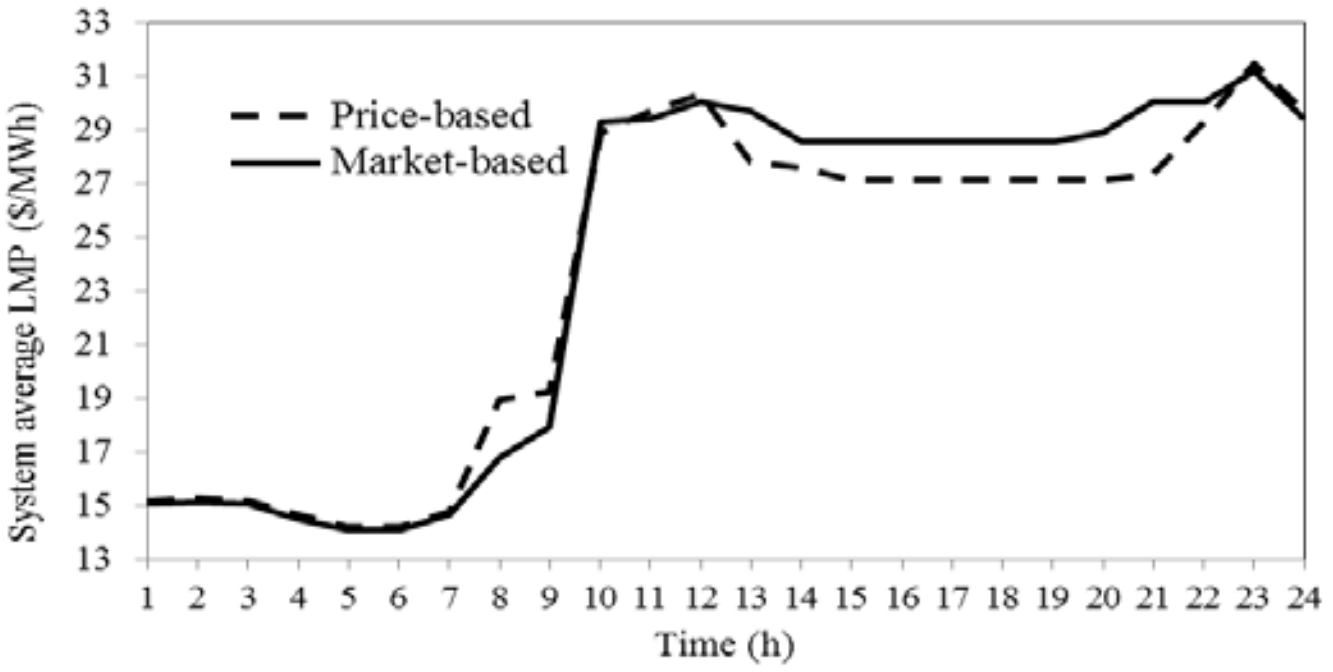

Fig. 7 Average LMP of 118-bus system

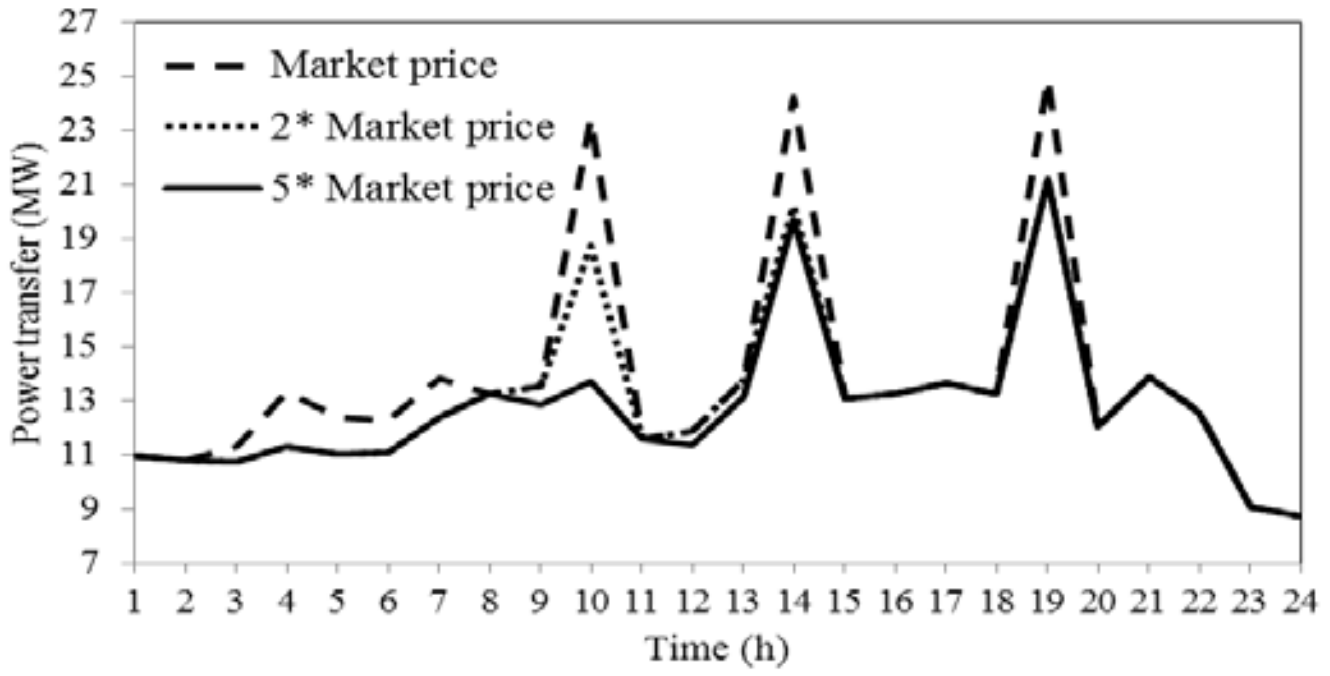

Fig. 8 Power transfer to microgrid 3 at different levels of deviation penalties.

To demonstrate the viability of the proposed deviation reduction method and ensuring that microgrid will follow the DMO assigned power transfers, the impact of the power transfer deviation penalty is further studied. Fig. 8 depicts the main grid power transfer to a selected microgrid (Microgrid 3) at different levels of deviation penalties. It is assumed that the forecasted microgrid load increases twofold at hours 10, 14 and 19.

Penalties equal to the market price, two times the market price, and five times the market price are considered. The cost of power transfer deviations are respectively calculated as $2,053, $1,777 and $3,170. As the penalty increases, the amount of deviation from assigned power decreases but the deviation cost does not change linearly. The microgrid total operation cost will however decrease. Thus it can be seen that higher penalties reduce the amount of deviation to reach the desired values. However, when the penalty becomes too high (comparable to the VOLL) microgrids may prefer to curtail some loads rather than paying for the penalty in purchasing power from the main grid.

In order to further show the impact of power deviation penalty on the scheduling solutions, two scenarios are considered; in the first scenario the microgrid is scheduled based on a price-based scheme after receiving the prices determined by the DMO; in the second scenario the absolute value of the power deviation, instead of only the positive deviation, is penalized. The microgrid operation cost reduces to $47,380 using price-based scheduling, as the microgrid reduces the power purchase from the main grid at the peak hours and uses its own resources that become price competitive at those times. When the absolute value of the deviations is penalized, the total microgrid operation cost rises to $50,539, as microgrid is obligated to closely follow the scheduled power transfer and hence would reduce generation of some its resources to purchase more power from the main grid, which results in a higher operation cost. This shows that penalizing power deviation is key to ensuring certainty in the power scheduled by the DMO. Penalizing the absolute value of power deviation can increase the microgrid operation cost, even if the power deviation would result in a surplus of power which is manageable by the system operator. The decision to penalize only the positive deviations or the absolute deviation should be made by the distribution system operator based on the probable congestion scenarios.

## VI. Conclusion

A market-based microgrid optimal scheduling model was proposed in this paper. The model was compared with the commonly used price-based scheduling model to show differences and exhibit merits. The market-based scheduling was performed by utilizing the DMO which would facilitate the market operations in the distribution system and act as an intermediate level between customers and the ISO. Simulations were performed using CPLEX and the obtained results were studied to show how microgrids can be optimally scheduled while taking distribution market decisions into account. Results showed that the market-based scheduling has the potential to lower the microgrid operation cost, lower LMPs at the microgrid-connected buses, and further support system operation by eliminating the net load uncertainty.

**Sina Parhizi** (S'13) received the B.S. degree in electrical engineering from the Sharif University of Technology, Tehran, Iran, in 2011 and the M.S. degree in electrical engineering from North Carolina State University, Raleigh, in 2013. He is pursuing the Ph.D. degree at the University of Denver in the Department of Electrical and Computer Engineering. His research interests include economics of electricity systems, microgrids, and distribution markets.

**Amin Khodaei** (SM'14) received the Ph.D. degree in electrical engineering from the Illinois Institute of Technology (IIT), Chicago, IL, USA, in 2010.

He was a visiting faculty (2010–2012) in the Robert W. Galvin Center for Electricity Innovation at IIT. He joined the University of Denver, Denver, CO, USA, in 2013 as an Assistant Professor. His research interests include power system operation, planning, computational economics, and smart electricity grids.

**Mohammad Shahidehpour** (F'01) is the Bodine Chair Professor and a director of the Robert W. Galvin Center for Electricity Innovation at the Illinois Institute of Technology, Chicago, IL, USA. He is a Research Professor at King Abdulaziz University, Jeddah, Saudi Arabia, and an Honorary Professor with North China Electric Power University, Beijing, China, and the Sharif University, Tehran, Iran.

Dr. Shahidehpour is the recipient of the Honorary Doctorate from the Polytechnic University of Bucharest, Bucharest, Romania, and the 2012 IEEE Power and Energy Society Outstanding Power Engineering Educator Award.